\documentclass[preprint,aps,prl,groupedaddress]{revtex4}

\usepackage{graphicx}
\usepackage{dcolumn}
\usepackage{bm}
\usepackage[T1]{fontenc}
\usepackage{color}

\usepackage[latin1]{inputenc}
\usepackage[T1]{fontenc}
\usepackage[francais]{babel}

\begin{document}
\title{Magneto-transport Subbands Spectroscopy in InAs Nanowires}
\author{Florian Vigneau $^{1}$}
\author{Vladimir Prudkovkiy $^{1}$}
\author{Ivan Duchemin $^{2}$}
\author{Walter Escoffier $^{1}$}
\author{Philippe Caroff $^{3,4}$}
\author{Yann-Michel Niquet $^{2}$}
\author{Renaud Leturcq $^{3}$}
\author{Michel Goiran $^{1}$}
\author{Bertrand Raquet $^{1}$}
\affiliation{$^{1}$Laboratoire National des Champs Magn\'{e}tiques
Intenses, INSA UPS CNRS, UPR 3228, Universit\'{e} de Toulouse, 143
av. de Rangueil, 31400 Toulouse, France}
\affiliation{$^{2}$ L-Sim, SP2M, UMR-E CEA/UJF-Grenoble 1, INAC, 17 rue des Martyrs, Grenoble, France }
\affiliation{$^{3}$Institute of Electronics Microelectronics and Nanotechnology,
CNRS-UMR 8520, ISEN Department, Avenue Poincar\'{e}, CS 60069, 59652 Villeneuve d'Ascq Cedex, France }
\affiliation{$^{4}$Department of Electronic Materials Engineering, Research School of Physics and Engineering,
The Australian National University, Canberra, ACT 0200, Australia}
\date{\today}

\begin{abstract}
We report on magneto-transport measurements in InAs nanowires under large magnetic field (up to 55T), providing a direct spectroscopy of the 1D electronic band structure. Large modulations of the magneto-conductance mediated by an accurate control of the Fermi energy reveal the Landau fragmentation, carrying the fingerprints of the confined InAs material. Our numerical simulations of the magnetic band structure consistently support the experimental results and reveal key parameters of the electronic confinement.
\end{abstract}

\maketitle

{\it Introduction}.- Semiconducting nanowires (sc-NW) represent a particular class of nano-objets with a broad range of potential applications in nano-electronics and optoelectronics ~\cite{Yangetplus, Dayeh1}: their aspect ratio facilitates their processing and combines well with the possibility of band structure tailoring and carrier doping. In particular, small band gap III-V sc-NWs such as InAs present key characteristics, deriving from their wide Bohr radius together with a strong spin-orbit interaction and a large Lande factor. The resulting unusual 1D electronic band structures constitute the foundation for spin and charge control in future nanodevices ~\cite{Bringer, Streda}. Recently, they have been exploited in the field of Majorana Fermions quest~\cite{Mourik}.

However, tailoring a reduced number conducting channels in 1D devices remains a challenge. It requires an electronic confinement that comes with a severe drop of the mobility ~\cite{Ford}. Despite tremendous effort, backscattering on defects is enhanced for the narrowest wires, concealing the 1D electronic properties ~\cite{Ford, Dayeh2}. Only recently, the first signatures of quasi-1D subbands in InAs NWs have been observed as steps in the conductance ~\cite{Ford2}. However, an exhaustive direct spectroscopy of the confined states in individual NWs remains unachieved.

Here, our strategy consists in playing with InAs NW based transistors in the open quantum dot regime and under extremely large magnetic field. High magnetic fields are required for a full spin and orbital degeneracy lifting once the magnetic confinement overcomes the electronic one ~\cite{Lassagne}. For a perpendicular magnetic field, the 1D electronic band structure evolves into magneto-electric subbands with a flattening of the dispersion curves, the onset of conducting chiral edge states and a Zeeman splitting together with an up-shift of the subbands energies accompanying the Landau diamagnetism ~\cite{Bogachek}. These magnetic states modify the conductance in a complex manner, following the depletion of the 1D channels and their degeneracy lifting in the quantum Hall regime~\cite{Bogachek, Royo}.

In this Letter, we give evidence of the 1D band structure of InAs NWs on the two-probe conductance mediated by the carrier density and a perpendicular magnetic field. Large conductance modulations reveal the magnetic field dependence of the 1D conducting states and the spin and orbital degeneracy. Under large magnetic field and/or at low carrier concentration, a full magnetic depletion of the NW is also achieved, inducing a turning off the conductance. Our results are consistently supported by numerical simulations of the magnetic band structure, revealing the key parameters of the electronic confinement in InAs.

InAs NWs with diameters (D) of $30 \pm 5nm$ are synthesized by gold-assisted gas-source molecular beam epitaxy on InP(111)B substrates ~\cite{Rolland}. High resolution transmission electron microscopy reveals a pure wurtzite crystal structure, virtually free of any extended structural defects~\cite{Xu}. The NWs are mechanically broken and deposited on a degenerately $n$-doped Si/SiO$_2$(225 nm) wafer. Contacts are defined along the NW using electron beam lithography. The oxide on the contact area is etched in a (NH$_4$)$_2$S$_x$ solution just before the evaporation of Ti/Au (10/150 nm) electrodes. The conductance of the InAs NW FETs versus back-gate voltage and pulsed magnetic field is measured from 300K down to 2K, using the standard lock-in technics in the low bias voltage regime ($eV_b<kT$) and under controlled atmosphere. Several devices (>10) have been fully characterized ensuring the robustness of our findings. In what follows, we mainly present extensive results for one device defined by a 225nm source-drain distance and $D=31nm$, having the hallmarks of the overall samples.

\begin{figure} [!ht]
\begin{center}
\includegraphics[bb=15 15 273 223, width=0.9\linewidth]{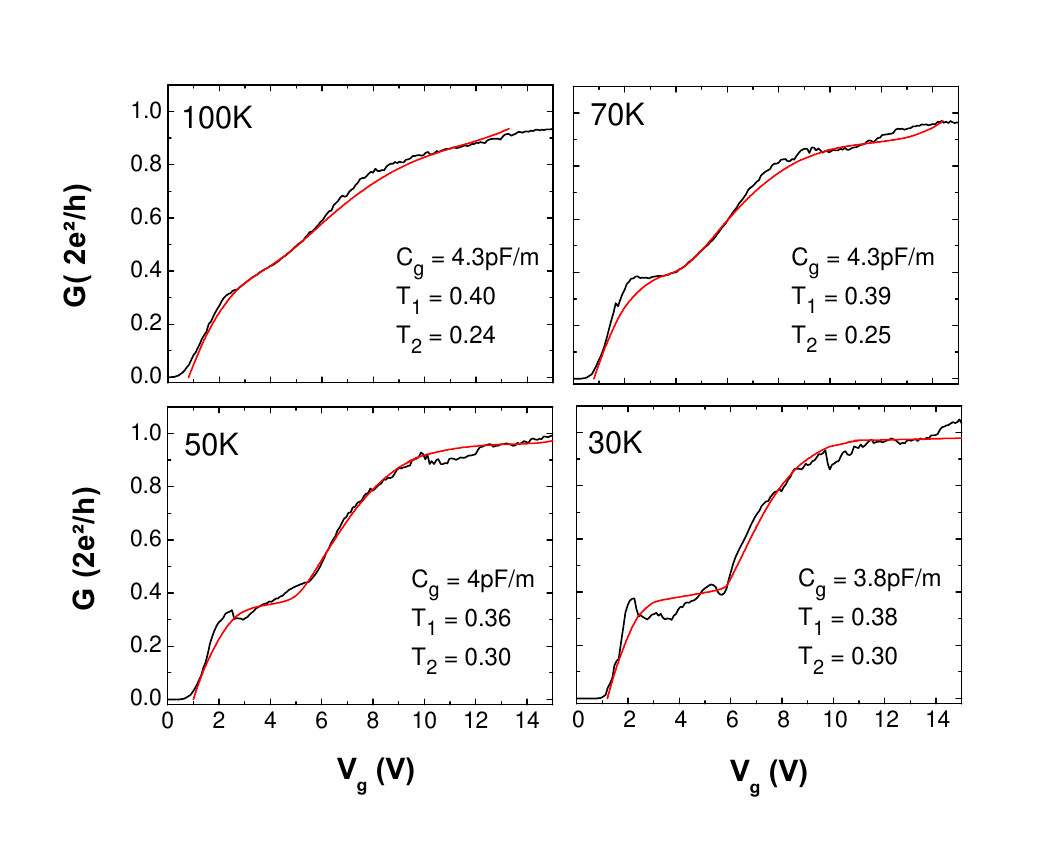}
\caption{(color online) Conductance versus $V_g$ measured at different temperatures on a 31nm diameter InAs NW device (black curves). Red curves are the simulated conductance following the Landauer formula. The electrostatic coupling and the transmission coefficients of the first two subbands are adjustable parameters.} \label{fig1}
\end{center}
\end{figure}

\begin{figure} [!ht]
\begin{center}
\includegraphics[bb=13 12 229 144, width=1\linewidth]{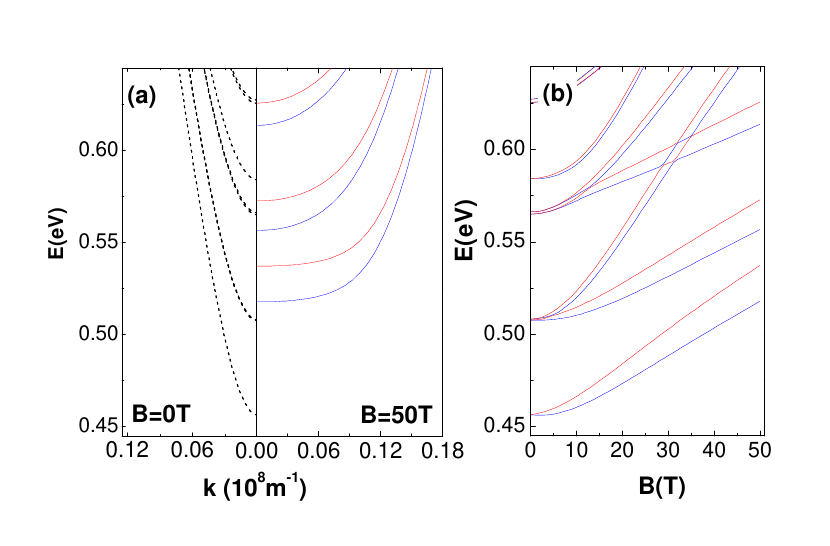}
\caption{(color online) (a) Simulated $E(k)$ dispersion curves along the NW axis for a 31nm diameter InAs NW, at 0T (black dot lines) and at 50T (red and blue lines for up and down spin states, respectively). (b) The corresponding magneto-electric subbands $E_i(B)$ from zero to 50T.} \label{fig2}
\end{center}
\end{figure}

The transfer characteristics $G(V_g)$ are presented in Fig.1 for selected temperatures (black lines). The conductance for a given carrier density is almost temperature independent while clear plateaus develop below 100K. Below 30K (not shown here), Universal Conductance Fluctuations (UCF) conceal the step-like behavior of the conductance. The presence of plateaus constitutes a straightforward signature of successive 1D-channels in a (quasi)-ballistic regime. Using the Landauer approach, the conductance is expressed by $G = 2e^2/h \sum \int -T_i(E)\partial f(E)/\partial E dE$, where $f(E)$ is the Fermi-Dirac distribution. The summation includes the 1D modes ($\emph{i}$) below $E_F$ and $T_i(E)$ refers to their transmission coefficient. The band structure of a 31nm diameter defect free InAs NW is calculated using the $sp^3d^5s^*$ tight-binding model~\cite{Jancu}. The dispersion curves are plotted in Fig.2a, left. Note that the second subband is two-fold degenerated as a consequence of the cylindrical symmetry. The relation between $E_F$ and $V_g$ voltage relies on the gating efficiency of the InAs device ($C_g$) and the integrated density of states. The modeling of the conductance with $T_i$ and $C_g$ as adjustable parameters (red curves, Fig.1) consistently reproduces the experimental results for all the temperatures. The transmission coefficients we extract per spin channel for the two first subbands are roughly temperature independent and equal to $T_{1,(2)} = 0.38(0.28) \pm 0.02$. These values are similar to those already reported on similar InAs NW devices~\cite{Ford2}. Assuming independent scattering events between the 1D channels and a negligible contact resistance checked by four-probes measurements (not shown here), we deduce the electronic mean free path $l_{1(2)} \approx 140 (90) nm$ for the first (second) subband, confirming the quasi-ballistic regime. The modeling of the conductance gives also an estimate of the geometrical capacitance. The knowledge of $C_g$ remains a key issue for an accurate determination of the field effect mobility ~\cite{Ford3}. External parameters like the metallic screening inherent to the contacts or the presence of surface states around the wire make inapplicable the routinely used metallic cylinder- plane capacitive model~\cite{Dayeh3, Khanal}. Here, we deduce an effective coupling of $4.2\pm 0.2 pF/m$, smaller than the simulated one ($13.7pF/m$), taking into account the screening of the metallic electrodes~\cite{Renaud2}, but modeling the nanowire as a metal. Our $G(V_g)$ analysis finally validates the sequence of the calculated conduction subbands minima and provides an accurate correspondence between  $E_F$ and $V_g$.

\begin{figure} [!ht]
\begin{center}
\includegraphics[bb=14 10 208 193, width=1\linewidth]{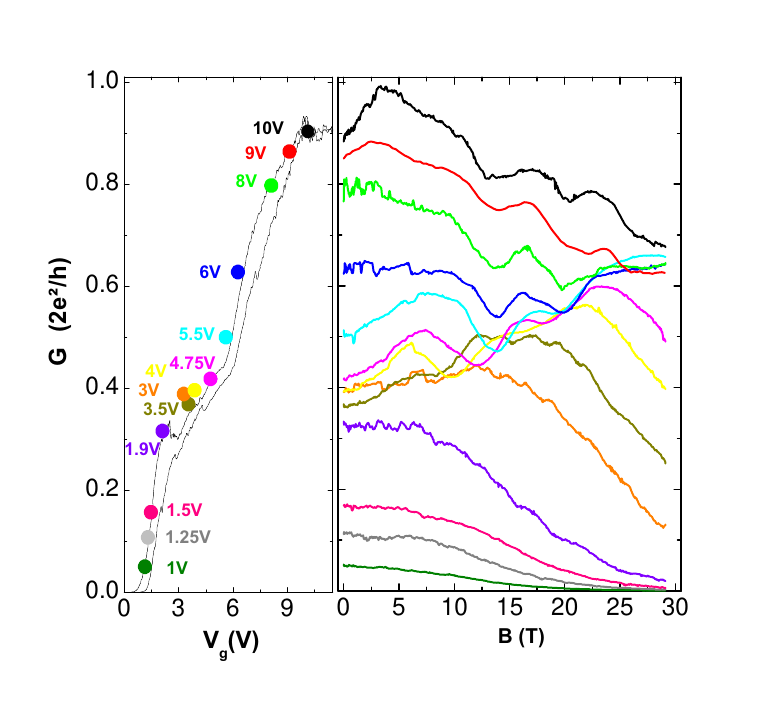}
\caption{(color online) Perpendicular magneto-conductance measured on the 31nm InAs NW device at 50K, for selected back-gate voltages (right panel), following the color marks on the $G(V_g)$ plot (left panel).} \label{fig3}
\end{center}
\end{figure}

Next, a magnetic field is applied perpendicular to the InAs NW and the magneto-conductance (MC) is measured for different carrier densities, following the color marks on the $G(V_g)$ plot (Fig.3). Experiments are performed at 50K to get rid off the UCF. Consequently, the diffusion coefficient driven by quantum interferences becomes mostly magnetic field independent; only the field induced changes of the electronic density of states affect the conductance. Large conductance modulations develop as a function of the magnetic field and depend on the number of 1D-channels below $E_F$. We schematically define three energy windows exhibiting distinct MC behavior : $\emph{(i)}$ Between the threshold voltage and the first conductance subband ($1V \leq V_g \leq 3V$), a monotonous decrease of the conductance is observed with a shift toward high magnetic field when the carrier density is increased. $\emph{(ii)}$ Below and in the vicinity of the second subband ($3.5V \leq V_g \leq 6V$), an oscillatory behavior of the conductance develops and consistently shifts to higher field when $E_F$ increases. These oscillations go along with an overall increase of the conductance followed by a drop of conductance at high field. $\emph{(iii)}$ Well inside the second subband ($6V < V_g$), the oscillatory behavior persists, accompanying an overall decrease of the conductance. Interestingly, several MC curves, with $ 5V \leq V_g \leq 9V$, merge together above 25T, suggesting a similar magnetic conducting state.

To rationalize these features, we first simulate the magnetic field dependence of the 1D $E_i(k)$ dispersion and their energy minima, $E_i(B)$ (Fig.2). Under 50T, the magnetic length (3.5nm) becomes much smaller than the NW radius. The electronic structure fragments into an unusual Landau spectrum, carrying the fingerprints of the confined InAs material. We clearly observe the flattening of the $E(k)$ curves at the $\Gamma$ point, reflecting the degenerate bulk Landau states, and the formation of the dispersive edge states when $k{l_B}^2$ approaches the flanks of the wire (Fig.2a right, red (blue) lines for the up (down) spin subbands). The $E_i(B)$ curves reveal an overall increase driven by the Landau diamagnetism and a strong spin and orbital degeneracy lifting in presence of a large $\mid g* \mid$ factor. At this stage, the Landauer formalism to model the $G(B)$ curves would result in step-like decrease of the conductance, following the magnetic depletion of the subbands. Our experimental results are more complex. For a deeper understanding, we compare the MC curves with the predicted magneto-electronic subbands and the field dependence of the Fermi energy $E_F(B)$, to accommodate the remaining available states below $E_F$. We assume a constant charge carrier density imposed by $V_g$.

\begin{figure} [!ht]
\begin{center}
\includegraphics[bb=14 10 208 193, width=1\linewidth]{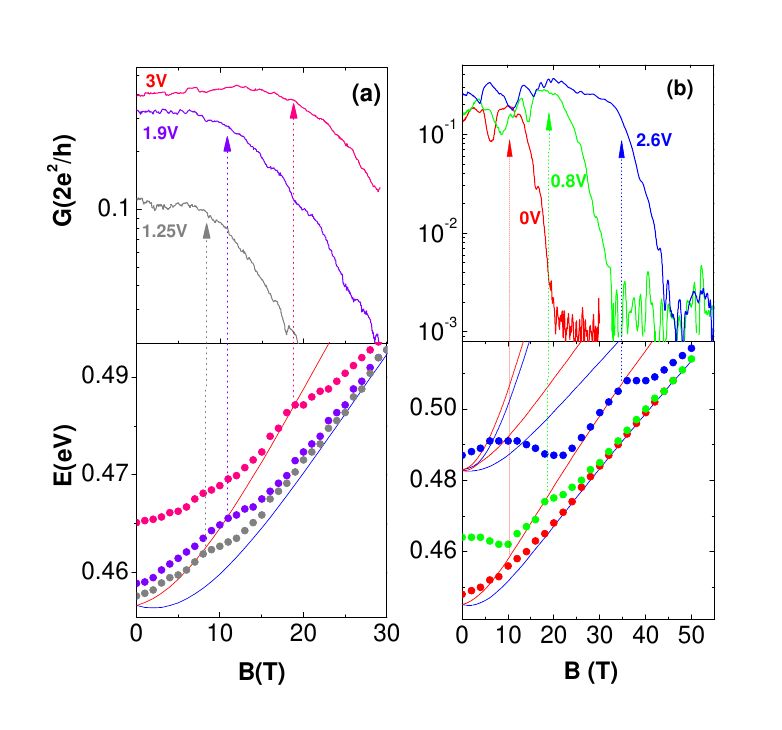}
\caption{(color online) (a) Zoom on the MC in the low doping regime measured on the 31nm diameter InAs NW device at 50K (top panel). Simulated magneto-electronic subbands and the corresponding $E_F(B)$ (low panel). The arrows indicate the crossing of $E_F$ with the subbands, concomitant with the conductance decrease. (b) MC measurements at 4K on a 33nm InAs NW device for selected gate voltage, revealing the switch-off of the conductance (top panel).  The corresponding $E_F(B)$ crossing the magneto-electric subbands demonstrate the magnetic depopulation (low panel).} \label{fig4}
\end{center}
\end{figure}

Fig.4a presents the analysis of the low doped regime $\emph{(i)}$. On the top panel, we plot the MC curves for $V_g$ equal 1.25, 1.9 and 3V. The corresponding $E_F(B)$ are superimposed to the magnetic field dependence of the first subband. We clearly observe an increase of $E_F$ with a slight pinning when crossing the bottom of the spin degeneracy lifted subband. The beginning of the conductance decrease coincides with the loss of the spin-up channel (marked by arrows). At 4K (Fig.4b, top panel), the drop of conductance, shown on an other device with a comparable diameter (33nm), is even more spectacular. It goes along with the pinning of $E_F$ on the lowest level (Fig.4b, low panel). Indeed, we assign the magnetic quench of the conductance to a complete electronic depletion of the wire. When $E_F$ approaches the bottom of the lowest subband, the charge carrier density can not be considered constant anymore. The quantum capacitance starts to be smaller than the geometrical capacitance in series, inducing to a strong decrease of the effective coupling and a depletion of the number of free electrons. This is in apparent contradiction with our assumption of a constant carrier density. However, we numerically checked that the quantum capacitance only plays a role at the bottom of the lowest subband which means that our simulated $E_F(B)$ remain valid until the high field drop of conductance.

\begin{figure} [!ht]
\begin{center}
\includegraphics[bb=9 12 205 193, width=1\linewidth]{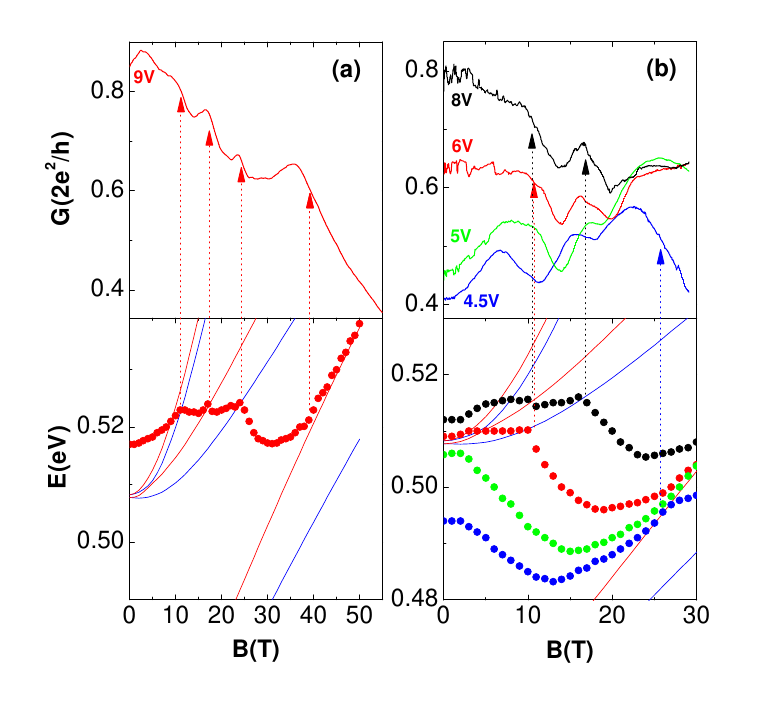}
\caption{(color online) (a) Zoom on the MC in the high doping regime (9V) measured on the 31nm diameter InAs NW device (top panel). The arrows indicate the successive drop of conductance unveiling the loss of conducting channels and the spin and orbital degeneracy lifting of the second subband (low panel). (b) Zoom in the intermediate doping regime (top panel). Only few structures of the $G(B)$ curves, marked by arrows, can be explained by the crossing of $E_F$ with magneto-electric subbands (low panel).} \label{Fig5}
\end{center}
\end{figure}

In the high doping regime $\emph{(iii)}$, when $E_F$ is well inside the second subband, the MC exhibits several bumps and step-like decreases (Fig.5a, top panel). The analysis of the magneto-electric subbands and their crossing of $E_F(B)$ (Fig.5a, low panel) reveals that all the successive drops of conductance are explained by the subband depopulations. An almost complete degeneracy lifting of the fourfold degenerate second subband is also resolved. Interestingly, the successive decreases are not quantized in unit of  $e^2/h$ and they are systematically preceded by an increase of the conductance. When $E_F$ lies between two magneto-electric sub-levels, the onset of chiral channels at the flanks induces a backscattering probability decrease. However, the magnetic depopulation occurs well before a complete spatial separation of the k+ and k- propagative states, preventing any quantized value of the conductance.

In the intermediate regime $\emph{(ii)}$, below and in the vicinity of the second subband, the behavior of the MC seems more complex (Fig.5b). At 8V, the first and the second drop of conductance, around 10T and 17T (marked by the black arrows), corresponds to the loss of the three first conducting channels and the fourth one of the second subband, respectively. At 6V, the first drop of conductance, occurring at 10T, is assigned to the crossing of the lowest sub-level of the second subband with no signature of a degeneracy lifting (marked by red arrow). The second drop of conductance, around 18T, does not find a straightforward explanation. The situation becomes even more puzzling when decreasing the doping level. At 4.5 and 5V, $E_F$ lies within 15meV below the second subband. An overall $\emph{increase}$ of the conductance with some large modulations develops. These modulations are similar to those observed at higher gate voltage, except that they occur at slightly lower magnetic fields. Here, we can not involve a direct crossing of $E_F$ with the magneto-electric subbands, except for the high field drop of conductance at 25T for $V_g=4.5V$ (marked by blue arrow). On the other hand, one should note that the rather high experimental temperature ($50K$) likely induces a fractional thermal occupation of the subbands lying few $kT$ (roughly $15meV$) above $E_F$. The first drop of conductance at 8 and 11T, for $V_g$ equals 4.5 et 5V, might be a consequence of the depopulation of the very few electrons thermally activated into the second subband. However, the well defined and reproducible shoulder we observe between 15T and 20T for $V_g$ in the range of $4-5.5V$ remain unexplained (Fig.3 and 5b).

In the following, we bring additional evidence of the magneto-fingerprints of the band structure. First, at low field, before the Landau regime (below 5T), we note that the MC successively varies from a flat signal, for $V_g \sim 1.5-3V$ and $6-8V$, to a large gain of conductance, for $V_g \sim 4-5.5V$ and $9-10V$ (Fig.3). The flat MC signal occurs when $E_F$ lies within the van Hove singularities (around $1.5 \pm 0.5V$ and $6\pm0.5V$). The corresponding k-vectors are strongly reduced and the resulting Lorentz force on the electronic trajectory is poorly effective. When $E_F$ is located well inside a conducting subband, much larger k-vectors are involved. The gain of velocity favors the classical force and the formation of skipping orbits at the flanks, responsible for the positive low field MC. This picture also explains the overall increase of G(B) for $V_g \sim 4-5.5V$. When $E_F$ is well above the highest occupied subband, here the first subband, the number of conducting channels remains constant over a wide magnetic field range and the magneto-conductance is mainly driven by the gradual onset of spatially separated chiral currents, responsible for a gain of conductance. Additionally, we previously mentioned that several MC curves merge together around 25T independently of the doping level, for $V_g$ varying from 5 to 9V. Fig.5 reveals that this is supported by a similar magnetic state when only the two lowest subbands carry the current in the high field regime.

We finally state that the major magneto-conductance features find natural explanations in the field dependence of the electronic band structure. The evidence of spin and orbital degeneracy lifting also constitutes an efficient tool to estimate the Zeeman energy and the $g^*$ factor. The first drop of conductance occurring at roughly $B \sim 10T$ for the curves at 6 and 8V coincides with two distinct $E_F$ ($E_{F,6V}\sim 509\pm 2meV$ and $E_{F,8V} \sim 515 \pm 2meV$) crossing the two sub-levels with similar orbital momentum but opposite spin states. We deduce the effective Lande factor for the second subband ($\mid g^*\mid = (E_{F,8V}-E_{F,6V})/\mu_BB)$, of the order of $8.4\pm 0.4$. Such a value, obtained in the open quantum dot regime, is significantly smaller than the InAs bulk value ($\mid g^* \mid\approx 15$), due to the transverse confinement ~\cite{Kiselev, Csonka}.

{\it Conclusion}.- By combining clean InAs NWs FET and large magnetic field environment, we bring a new insight into the magneto-transport spectroscopy of the 1D modes in the open quantum dot regime. A direct comparison with simulations reveal the magneto-electric subbands and their major impacts on the conductance, from which key parameters like the $g^*$ factor can be deduced. This opens news routes toward a magnetic tailoring of the 1D band structure in sc-NWs.

{\it Acknowledgements}.- Part of this work is supported by EuroMagNET, contract n$^o$ 228043 and the ANR Project No. ANR-11-JS04-002-01. The authors thank Xavier Wallart for helpfull discussions.


\begin{thebibliography}{50}

\bibitem{Yangetplus}
P. Yang, R. Yan and M. Fardy, Nano Lett. $\textbf{10}$, 1529 (2010); N. S. Ramgir, Y. Yang and M. Zacharias, Small $\textbf{6}$, 1705 (2010); Y. Li, F. Qian, J. Xiang and Ch.M. Lieber, Materials Today $\textbf{9}$, 18 (2006).

\bibitem{Dayeh1}
S. A. Dayeh, Semicond. Sci. Technol. $\textbf{25}$, 024004 (2010).

\bibitem{Bringer}
A. Bringer and Th. Schäpers, Phys. Rev. B $\textbf{83}$, 115305 (2011).

\bibitem{Streda}
P. Streda and P. Seba, Phys. Rev. Lett. $\textbf{90}$, 256601 (2003).

\bibitem{Mourik}
V. Mourik, K. Zuo, S.M. Frolov, S.R. Plissard, E.P.A.M. Bakkers and L.P. Kouwenhoven, Science $\textbf{336}$, 1003 (2012); M. T. Deng, C. L. Yu, G. Y. Huang, M. Larsson, P. Caroff, and H. Q. Xu, Nano Letters, \textbf{12}, 6414 (2012);
A. Das, Y. Ronen, Y. Most, Y. Oreg, M. Heiblum, and H. Shtrikman, Nature Physics, \textbf{8}, 887 (2012).


\bibitem{Ford}
A. C. Ford, J. C. Ho, Yu-Lun Chueh, Yu-Chih Tseng, Zhiyong Fan, Jing Guo, J. Bokor, A. Javey, NanoLett. $\textbf{9}$, 360 (2009); N. Gupta, Y. Song, G. W. Holloway, U. Sinha, C. M. Haapamaki, R. R. LaPierre and J. Baugh, Nanotechnolgy $\textbf{24}$, 225202 (2013).

\bibitem{Dayeh2}
S.A. Dayeh, C. Soci, P.K.L. Yu, E.T. Yu and D. Wang, J. Vacc. Scien. And Tech. B, $\textbf{25}$, 1432 (2007);

\bibitem{Ford2}
A. C. Ford, S. B. Kumar, R. Kapadia, J. Guo and A. Javey, Nano Lett. $\textbf{12}$,1340 (2012); S. Chuang, Q. Gao, R. Kapadia, A. C. Ford, J. Guo and A. Javey, Nano Lett. $\textbf{13}$, 555 (2013).

\bibitem{Lassagne}
S. Nanot, R. Avriller, W. Escoffier, J-M Broto, S. Roche and B. Raquet, Phys. Rev. Lett. $\textbf{103}$, 256801 (2009).

\bibitem{Bogachek}
E. N. Bogachek, A. G. Scherbakov and U. Landman, Phys. Rev. B $\textbf{53}$, (R)13246 (1996); J. Knobbe and Th. Schapers, Phys. Rev. B $\textbf{71}$, 035311 (2005); Y. Tsrkovnyak and B. I. Halperin, Phys. Rev. B $\textbf{74}$, 245327 (2006).

\bibitem{Royo}
M. Royo, A. Bertoni and G. Goldoni, Phys. Rev. B $\textbf{87}$, 115316 (2013).


\bibitem{Rolland}
C. Rolland, P. Caroff, C. Coinon, X. Wallart, and R. Leturcq, Appl. Phys. Lett. \textbf{102}, 223105 (2013).

\bibitem{Xu}
T. Xu, K. A Dick, S. Plissard, T. H. Nguyen, Y. Makoudi, M. Berthe, J-P Nys, X. Wallart, B. Grandidier and P. Caroff, Nanotechnology \textbf{23}, 095702 (2012).

\bibitem{Jancu}
J.M. Jancu, R. Scholz, F. Beltram and F. Bassani, Phys. Rev. B \textbf{57}, 6493 (1998); Y. M. Niquet, A. Lherbier, N. H. Quang, M. V. Fernandez-Serra, X. Blase and C. Delerue, Phys. Rev. B \textbf{73}, 165319 (2006).

\bibitem{Ford3}
A. C. Ford, J. C. Ho, Y-L. Chueh, Y-C. Tseng, Z. Fan, J. Guo, J. Bokor and A. Javey, Nano Lett. $\textbf{9}$, 360 (2009).

\bibitem{Dayeh3}
S. A. Dayeh, C. Soci, P. K. L. Yu, E. T. Yu and D. Wang, J. Vac. Sci. Technol. B $\textbf{25}$, 1432 (2007).

\bibitem{Khanal}
D. R. Khanal and J. Wu, Nano Lett. $\textbf{7}$, 2778 (2007).

\bibitem{Renaud2}
The capacitance of the device including the contacts was calculated by a 3D electrostatic simulation using COMSOL MULTIPHYSICS.

\bibitem{Kiselev}
A. A. Kiselev and E. L. Ivchenko, Phys. Rev. B \textbf{58}, 16358 (1998).

\bibitem{Csonka}
S. Csonka, L. Hofstetter, F. Freitag, S. Oberholzer and C. Schonenberger, Nano Lett. $\textbf{8}$, 3932 (2008).




\end{thebibliography}
\end{document}